\documentclass[%
 reprint,
superscriptaddress,
 amsmath,amssymb,
 aps,
]{revtex4-2}

\usepackage{graphicx}% Include figure files
\usepackage{dcolumn}% Align table columns on decimal point
\usepackage{bm}% bold math

\usepackage{physics}
\usepackage{mhchem}
\usepackage{natbib}

\usepackage{color}

\usepackage[normalem]{ulem}

\begin{document}

\title{A new type of skin thickness in high-spin isomers
to constrain equation of state of spin-polarized nuclear matter}

\author{Toi Tachibana}
\affiliation{Department of Physics, Kyoto University, Kyoto 606-8502, Japan}

\author{Kouichi Hagino}

\affiliation{Department of Physics, Kyoto University, Kyoto 606-8502, Japan}
\affiliation{Institute for Liberal Arts and Sciences, Kyoto University, Kyoto 606-8501, Japan}
\affiliation{ 
RIKEN Nishina Center for Accelerator-based Science, RIKEN, Wako 351-0198, Japan}

\author{Kenichi Yoshida}
\affiliation{Research Center for Nuclear Physics, The University of Osaka, Ibaraki, 567-0047, Japan}
\affiliation{ 
RIKEN Nishina Center for Accelerator-based Science, RIKEN, Wako 351-0198, Japan}
\affiliation{Center for Computational Sciences, University of Tsukuba, Tsukuba, Ibaraki 305-8577, Japan}

\author{Qiang Zhao}
\affiliation{China Institute of Atomic Energy (CIAE), P. O. Box 275(18), Beijing 102413, China}
\affiliation{Research Center for Nuclear Physics, The University of Osaka, Ibaraki, 567-0047, Japan}

\date{\today}

\begin{abstract}
We explore experimental probes
for constraining the equation of state (EOS)
of spin-polarized nuclear matter,
where spins of nucleons are aligned along a particular direction.
For this purpose,
we calculate the $12^+$ isomeric state of the \ce{^{52}Fe} nucleus
with the relativistic point-coupling model. 
We argue that 
the spin skin thickness of this high-spin state, that is, the difference 
between the radius of the spin-up density and that of the spin-down density, 
strongly correlates with the spin slope parameter of the EOS. 
This is in analogy to the well known linear correlation
between the slope parameter of the symmetry energy
and the neutron skin thickness of finite nuclei.
We show that this correlation is retained even when one considers the difference of the radii 
between the $12^+$ and the ground states, even though the correlation is much weaker than in the 
case of the difference between the spin-up and the spin-down radii. 
We also discuss the mean-square charge radii of the $12^+$ state. We find that 
by taking the square of the radii the dependence on the spin slope parameter becomes much stronger, 
and thus it can serve as a good probe of the spin slope parameter given a high resolution of 
laser-spectroscopy experiments. 
\end{abstract}

\maketitle

%**************************************************
%        SECTION 1
%**************************************************

\section{INTRODUCTION}
\label{sec:intro}

The equation of state (EOS) of nuclear matter 
is fundamental to understanding various astrophysical processes, 
including the dynamics of supernovae 
and the inner structure of neutron stars 
\cite{Yamamoto_Togashi,Togashi_2017,Shen,Hempel_2012,Sumiyoshi_2005}. 
For instance, the EOS plays a decisive role in determining the mass-radius relation 
of neutron stars \cite{Oertel2017}. 
Because of this, the precise determination of the EOS 
through a unified theoretical framework applicable to 
both finite nuclei and infinite nuclear matter
has been one of major objectives in nuclear physics.
In particular, the EOS of asymmetric nuclear matter plays an essential role
in these astrophysical phenomena, 
given the extreme conditions present in compact objects, 
which are characterized by high isospin asymmetry 
and a wide range of nucleon density. 
This fact has motivated extensive theoretical studies
\cite{APR,Togashi_2013,Bompaci_1991,Lattimer_2013}
directed towards determining the EOS in the wide-density 
and neutron-rich regions.

It has been suggested that terrestrial nuclear experiments 
can constrain the symmetry energy near the saturation density, 
which is roughly defined as the energy difference between
symmetric nuclear matter and pure neutron matter.
A prominent example is the well-known linear correlation 
between the slope parameter, which is defined as the derivative
of the symmetry energy at the saturation density, 
and the neutron skin thickness of relatively neutron rich nuclei
such as \ce{^{208}Pb} and \ce{^{48}Ca} \cite{Roca-maza}.
This correlation has motivated numerous experiments
aiming at accurate measurements of the neutron skin thicknesses 
\cite{Zenihiro,Tamii2011,PREX,CREX,Carbone2010}.
See also Refs. \cite{Berman_1975,Tsang2012} for other types of measurements to experimentally 
explore the EOS of asymmetric nuclear matter.

While 
the total spin of nuclear matter is assumed to be zero in these previous studies, 
the EOS of spin-polarized matter, in which there is an asymmetry between spin-up 
and spin-down nucleons, 
has recently attracted significant interests. 
This was largely triggered by  
the observations of the blue kilonova ejecta associated 
with the GW170817 neutron star merger \cite{kilonova,GW170817,Evans}, 
that suggested the existence of a hypermassive neutron star remnant 
hosting an intense surface magnetic field on the order of $(1-3)\times10^{14}$ G 
\cite{Metzger_2018}. 
Despite this growing astrophysical relevance, 
almost no observable of terrestrial nuclear experiments 
has been considered, 
which can constrain the EOS
of spin-polarized matter. 

In Ref. \cite{Tachibana2025}, 
we studied the spin-polarized nuclear matter
by using the relativistic framework, 
which more naturally treats the spin degree of freedom of nucleons 
than non-relativistic frameworks.
In addition,
we investigated the spin slope parameter $L_s$,
which is defined \cite{Khoa_2020,Khoa_2022} in the same manner as for the 
slope parameter $L$ of spin-unpolarized nuclear matter. 
We showed that $L$ and $L_s$ for isoscalar polarization has a strong correlation. 
In this way, $L_s$ could be determined indirectly through the neutron skin thickness, which can be 
used to fix the value of $L$. 

In this paper, we explore an experimental probe which is more directly correlated with 
$L_s$. To this end,  
we calculate 
the $12^+$ isomeric state of the \ce{^{52}Fe} nucleus. 
The $12^+$ state of \ce{^{52}Fe} located at $6.96$ MeV \cite{Gadea2005} 
is known as a yrast trap isomer,
since it lies below the yrast $10^+$ state 
at $7.38$ MeV \cite{Ur1998}.
Because of this level inversion,
the E2 decay to the $10^+$ state is energetically forbidden,
which makes the $12^+$ state a long-lived isomer.
This state decays almost exclusively into excited states of \ce{^{52}Mn}
via Gamow-Teller transitions with a branching ratio of $99.98\%$,
and has a long half-life of $45.9$ s \cite{Geesaman1979}.
Because of this long half-life, 
it may be possible to measure precisely the radii of this 
$12^+$ isomeric state. 

In Ref. \cite{Liang2018}, 
Liang et al. calculated the transition strengths
of Gamow-Teller giant resonances (GTGR)
from the $12^+$ isomeric state of \ce{^{52}Fe}
to the excited states of \ce{^{52}Co}.  
They pointed out that 
the different Fermi energies between the spin-up and the spin-down nucleons in the isomeric state  
play a similar role in GTGR as those of neutrons and protons in the ground state of neutron-rich nuclei. 
In other words, 
the spin degree of freedom takes over the role played by isospin.
Considering the fact that the neutron skin thickness
reflects the difference in the Fermi surfaces between
the neutrons and protons of neutron-rich nuclei,
we shall 
introduce a spin "skin thickness"
of the $12^+$ isomeric state of the \ce{^{52}Fe} nucleus.

From the nuclear structure point of view,
shell-model calculations for \ce{^{52}Fe} with the model space of the full $pf$ shell describe 
the $12^+$ state as the configuration 
in which both protons and neutrons dominantly occupy the $f_{7/2}$ orbital
with their angular momenta coupled 
to the maximum value \cite{Ur1998,Gadea2005}.
Since the configuration of this state is well identified,
in this paper we perform a configuration-fixed self-consistent calculation
by using the PCF-PK1 parameter set \cite{PCF-PK1} 
for 
a relativistic point-coupling model
including the Fock terms. 
Because of the Fock terms, with this framework
one can treat the time-odd coupling terms 
which play a crucial role in spin-polarized matter
and high-spin isomeric states.
Using this framework, 
we shall investigate the correlation between the spin slope parameter
and the spin skin thickness of \ce{^{52}Fe}.

The paper is organized as follows.  
In Sec. \ref{se:PCF_Todd}, we formulate 
the relativistic point-coupling model. 
In Sec \ref{sec:52Fe12+},
we carry out the configuration-fixed self-consistent calculation
for the $12^+$ isomeric state of the \ce{^{52}Fe} nucleus. 
In Sec. \ref{sec:correlation},
we present the correlation between the spin slope parameter
and the spin skin thickness of the \ce{^{52}Fe} nucleus.
Finally, we summarize the paper in Sec. \ref{sec:summary}.

%**************************************************
%        SECTION 2
%**************************************************

\section{Relativistic Point-Coupling Model for high-spin isomers without time-reversal symmetry} 
\label{se:PCF_Todd}

In this paper, we employ the relativistic point-coupling model \cite{PCF-PK1,PC-F1,PC-PK1} 
in order to calculate the EOS of spin-polarized matter as well as 
the $12^+$ isomeric state of \ce{^{52}Fe}.
For this purpose, 
we start with a general form of the energy functional
and explain the treatment for systems 
in which the time-reversal symmetry is broken.

When the pairing correlation is not taken into account, 
the total energy functional in this model reads
\begin{align}
    E_{\text{tot}} = E_{\text{kin}} 
    + E_{\text{4f}} + E_{\text{der}} + E_{\text{em}}.
    \label{eq:e_tot}
\end{align}
Here, $E_{\text{kin}}$ is the kinetic energy term given by
\begin{align}
    E_{\text{kin}} = \int d^3\bm{r} \sum_{\alpha} \psi^{\dagger}_{\alpha}
    \left( -i\bm{\alpha}\cdot\nabla + M \beta \right) \psi_{\alpha},
\end{align}
where $\psi_{\alpha}(\bm{r})$ represents a single-particle wave function of nucleons
and $\alpha$ is the index for the single-particle basis.
$E_{\text{4f}}$ is the four-fermion point-coupling term expressed as
\begin{align}
    \label{eq:4f_energy}
    E_{\text{4f}} = & \frac{1}{2} \int d^3 \bm{r} 
    \left[ 
    \alpha_S \left(\sum_{\alpha}\bar{\psi}_{\alpha}
    \psi_{\alpha}\right)^2 
    + \alpha_{tS} \left(\sum_{\alpha}\bar{\psi}_{\alpha}
    \vec{\tau}\psi_{\alpha}\right)^2
    \right. \notag \\
    &
    + \alpha_V \left(\sum_{\alpha}\bar{\psi}_{\alpha}
    \gamma^{\mu}\psi_{\alpha}\right)^2 
    + \alpha_{tV} \left(\sum_{\alpha}\bar{\psi}_{\alpha}
    \gamma^{\mu}\vec{\tau}\psi_{\alpha}\right)^2 
    \notag \\
    &
    + \alpha_{PS} \left(\sum_{\alpha}\bar{\psi}_{\alpha}
    \gamma_5\psi_{\alpha}\right)^2 
    + \alpha_{tPS} \left(\sum_{\alpha}\bar{\psi}_{\alpha}
    \gamma_5\vec{\tau}\psi_{\alpha}\right)^2 
    \notag \\
    &
    + \alpha_{PV} \left(\sum_{\alpha}\bar{\psi}_{\alpha}
    \gamma_5\gamma^{\mu}\psi_{\alpha}\right)^2 
    + \alpha_{tPV} \left(\sum_{\alpha}\bar{\psi}_{\alpha}
    \gamma_5\gamma^{\mu}\vec{\tau}\psi_{\alpha}\right)^2 
    \notag \\
    & \left.
    + \alpha_T \left(\sum_{\alpha}\bar{\psi}_{\alpha}
    \sigma^{\mu\nu}\psi_{\alpha}\right)^2 
    + \alpha_{tT} \left(\sum_{\alpha}\bar{\psi}_{\alpha}
    \sigma^{\mu\nu}\vec{\tau}\psi_{\alpha}\right)^2 
    \right].
\end{align}
Here, the subscripts $S$, $V$, $PS$, $PV$ and $T$ represent the scalar, vector, 
pseudoscalar, pseudovector and tensor coupling terms, respectively, 
while the additional subscript \lq\lq$t$" refers to the isovector channel.
We denote this term as
\begin{align}
    E_{\text{4f}} = \frac{1}{2} \int d^3 \bm{r}
    \sum_{i} \alpha_i
    \left[ \sum_{\alpha} \bar{\psi}_{\alpha} 
    \left( \mathcal{O}\Gamma \right)_i \psi_{\alpha} \right]^2,
\end{align}
where the matrices $\mathcal{O}$ and $\Gamma$ represent 
the isospin matrix and the gamma matrix, respectively, with 
\begin{align}
    \mathcal{O}\in\{1,\tau_3\}, ~~~~
    \Gamma\in\{1,\gamma_5,\gamma^{\mu},\gamma_5\gamma^{\mu},\sigma^{\mu\nu}\}.
\end{align}
Each combination of $\mathcal{O}$ and $\Gamma$,  that is, $\left(\mathcal{O}\Gamma\right)_i$, corresponds to a coupling channel, $i\in\{S, tS, V, tV, PS, tPS, PV, tPV, T, tT\}$. 
$E_{\text{der}}$ in Eq. \eqref{eq:e_tot} is the derivative term given by 
\begin{align}
    E_{\text{der}} = - \frac{1}{2} \int d^3\bm{r} \delta_S 
    \left[ \nabla \left(\sum_{\alpha} \bar{\psi}_{\alpha}
    \psi_{\alpha} \right) \right]^2,
\end{align}
and $E_{\text{em}}$ is the electromagnetic term given as 
\begin{align}
    E_{\text{em}} &= \frac{1}{2} \frac{e^2}{4\pi} 
    \int \int d^3\bm{r} d^3\bm{r}'
    \left[ \sum_{\alpha} \bar{\psi}_{\alpha}(\bm{r}) 
    \gamma_{\mu} \frac{1-\tau_3}{2} \psi_{\alpha}(\bm{r}) 
    \right] \notag \\
    & \times \frac{1}{|\bm{r}-\bm{r}'|}
    \left[ \sum_{\beta} \bar{\psi}_{\beta}(\bm{r}') 
    \gamma^{\mu} \frac{1-\tau_3}{2} \psi_{\beta}(\bm{r}') 
    \right],
\end{align}
where the Coulomb exchange term is neglected for simplicity.

Although ten coupling constants are included
in Eq. \eqref{eq:4f_energy},
not all of them are independent.
The number of independent coupling-constants is five,
and we take $\alpha_S$, $\alpha_V$, $\alpha_{tS}$,
$\alpha_{tV}$ and $\alpha_T$ as those independent constants \cite{PCF-PK1}.
The other coupling constants are given by linear combinations
of the five independent coupling constants,
\begin{subequations}
\label{eq:rel_cc}
\begin{align}
    \alpha_{PS}&=\frac{1}{3}(-\alpha_S-6\alpha_{tS}-4\alpha_{V}+12\alpha_{tV}-12\alpha_T)
    \label{eq;alpha_PS}\\
    \alpha_{tPS}&=\frac{1}{9}(-4\alpha_S+8\alpha_{tS}+2\alpha_{V}-24\alpha_{tV}-12\alpha_T)
    \label{eq:alpha_tPS}\\
    \alpha_{PV}&=\frac{1}{3}(2\alpha_S+3\alpha_{tS}+2\alpha_{V}+3\alpha_{tV}+6\alpha_T)
    \label{eq:alpha_PV}\\
    \alpha_{tPV}&=\frac{1}{9}(2\alpha_S+3\alpha_{tS}+5\alpha_{V}-6\alpha_{tV}+6\alpha_T)
    \label{eq:alpha_tPV}\\
    \alpha_{tT}&=\frac{1}{18}(-\alpha_S+3\alpha_{tS}+2\alpha_{V}-6\alpha_{tV}+6\alpha_T).
    \label{eq:alpha_tT}
\end{align}
\end{subequations}

For open-shell nuclei, 
pairing correlations should be included in calculations.
In the Hartree-Fock-Bogoliubov (HFB) framework, 
the energy associated with a pairing interaction $V_{\text{pair}}$ is given by 
\begin{align}
    E_{\text{pair}} = \frac{1}{2}
    \Tr[\Delta\kappa],
\end{align}
where $\kappa$ represents the pairing density \cite{RingSchuck},
and $\Delta$ is the pairing field given by
\begin{align}
    \Delta_{\alpha\beta} = \frac{1}{2}
    \sum_{\gamma\delta} \bra{\alpha\beta} V_{\text{pair}}
    \ket{\gamma\delta} \kappa_{\gamma\delta}.
\end{align}
In the relativistic point-coupling model \cite{PCF-PK1},
the separable finite-range pairing interaction is employed 
with parameters adjusted to the pairing gap of the Gogny D1S interaction \cite{Tian2009}.  

The relativistic HFB equation
is obtained by taking the variation of the energy density functional
as
\begin{align}
    \left(\begin{array}{cc}
        \hat{h} - \lambda & \hat{\Delta} \\
        \hat{\Delta}^* & - \hat{h}^* + \lambda
    \end{array}\right)
    \left(\begin{array}{c}
        U_{\alpha} \\
        V_{\alpha}
    \end{array}\right)
    = E_{\alpha} \left(\begin{array}{c}
        U_{\alpha} \\
        V_{\alpha}
    \end{array}\right),
\end{align}
where $\hat{h}$ represents the single-particle Hamiltonian
expressed as
\begin{align}
    \label{eq:1p_ham_gene}
    \hat{h} = & -i\bm{\alpha}\cdot\nabla 
    + \left(M + \delta_S \nabla^2 \sum_{\beta}
    \bar{V}_{\beta}(\bm{r})V_{\beta}(\bm{r})
    \right) \beta
    \notag \\
    & + \sum_{i} \alpha_i \left[ \sum_{\beta} \bar{V}_{\beta}(\bm{r})
    \left(\mathcal{O}\Gamma\right)_i V_{\beta}(\bm{r}) \right]
    \beta \left(\mathcal{O}\Gamma\right)_i,
\end{align}
and $\lambda$ represents the Fermi energy.
For a system with time-reversal invariance,
such as the $0^+$ ground state of even-even nuclei,
only the scalar density, the time component of the vector current
and the $0i$ component of the tensor density ($i=1,2,3$) 
take non-zero values,
while the other components vanish.
We denote these non-vanishing components as
\begin{subequations}
\label{eq:densities_even}
\begin{align}
    \rho_S(\bm{r}) &\equiv \sum_{\alpha}
    \bar{V}_{\alpha}(\bm{r}) V_{\alpha}(\bm{r}) \\
    \rho_{tS}(\bm{r}) &\equiv \sum_{\alpha} \tau_3^{(\alpha)}
    \bar{V}_{\alpha}(\bm{r}) V_{\alpha}(\bm{r}) \\
    \rho_B(\bm{r}) &\equiv \sum_{\alpha}
    V^{\dagger}_{\alpha}(\bm{r}) V_{\alpha}(\bm{r}) 
    ~~ ( = j_V^0(\bm{r})) \\
    \rho_{tB}(\bm{r}) &\equiv \sum_{\alpha} \tau_3^{(\alpha)}
    V^{\dagger}_{\alpha}(\bm{r}) V_{\alpha}(\bm{r})
    ~~ ( = j_{tV}^0(\bm{r})) \\
    j_{Te}^i(\bm{r}) &\equiv \sum_{\alpha}
    \bar{V}_{\alpha}(\bm{r}) \sigma^{0i} V_{\alpha}(\bm{r})
    ~~ (i = 1,2,3) \\
    j_{tTe}^i(\bm{r}) &\equiv \sum_{\alpha} \tau_3^{(\alpha)}
    \bar{V}_{\alpha}(\bm{r}) \sigma^{0i} V_{\alpha}(\bm{r})
    ~~ (i = 1,2,3),
\end{align}
\end{subequations}
where the subscript $Te$ represents $0i$ components,
that is, time-even components of the tensor density.
Then, one can rewrite the single-particle Hamiltonian
\eqref{eq:1p_ham_gene} as
\begin{align}
    \hat{h} = & -i\bm{\alpha}\cdot\nabla + \beta M^*(\bm{r})
    + V^0_V(\bm{r}) - i \beta\bm{\alpha} \cdot \bm{V}_{Te}(\bm{r}),
\end{align}
where each potential and effective mass are given by
\begin{subequations}
\label{eq:potential_even}
\begin{align}
    M^*(\bm{r}) &= M + \alpha_S \rho_S(\bm{r}) + \tau_3 \alpha_{tS} \rho_{tS}(\bm{r})
    + \delta_S \nabla^2 \rho_S(\bm{r}) \\
    V^0_V(\bm{r}) &= \alpha_V \rho_B(\bm{r}) + \tau_3 \alpha_{tV} \rho_{tB}(\bm{r})
    + e \frac{1-\tau_3}{2} A^0(\bm{r}) \\
    \bm{V}_{Te}(\bm{r}) &= 2(\alpha_T \bm{j}_{Te}(\bm{r}) + \tau_3
    \alpha_{tT} \bm{j}_{tTe}(\bm{r})).
\end{align}
\end{subequations}
Notice that the tensor potential $\bm{V}_{Te}$ is multiplied
by a factor $2$ because of the antisymmetric property of
the tensor matrix $\sigma^{\mu\nu}$.

We focus on a high-spin state of nuclei, 
in which the time-reversal symmetry is broken.
Then, one has to consider not only the time-even potentials
given by Eq. \eqref{eq:potential_even}
but also the time-odd potentials, 
which vanish in a system with the time-reversal symmetry.
The additional non-vanishing terms are the spacial components of 
the vector and the pseudovector currents,
the pseudoscalar density and the $ij$ components of tensor density,
which are expressed as
\begin{subequations}
\label{eq:densities_odd}
\begin{align}
    j_{V}^i(\bm{r}) &\equiv \sum_{\alpha}
    \bar{V}_{\alpha}(\bm{r}) \gamma^{i} V_{\alpha}(\bm{r}) \\
    j_{tV}^i(\bm{r}) &\equiv \sum_{\alpha} \tau_3^{(\alpha)}
    \bar{V}_{\alpha}(\bm{r}) \gamma^{i} V_{\alpha}(\bm{r}) \\
    \rho_{PS}(\bm{r}) &\equiv i\sum_{\alpha}
    \bar{V}_{\alpha}(\bm{r}) \gamma_5 V_{\alpha}(\bm{r}) \\
    \rho_{tPS}(\bm{r}) &\equiv i\sum_{\alpha} \tau_3^{(\alpha)}
    \bar{V}_{\alpha}(\bm{r}) \gamma_5 V_{\alpha}(\bm{r}) \\
    j_{PV}^i(\bm{r}) &\equiv \sum_{\alpha}
    \bar{V}_{\alpha}(\bm{r}) \gamma^{i} \gamma_5 V_{\alpha}(\bm{r}) \\
    j_{tPV}^i(\bm{r}) &\equiv \sum_{\alpha} \tau_3^{(\alpha)}
    \bar{V}_{\alpha}(\bm{r}) \gamma^{i} \gamma_5 V_{\alpha}(\bm{r}) \\
    j_{To}^i(\bm{r}) &\equiv \sum_{\alpha} \frac{1}{2} \epsilon^{ijk}
    \bar{V}_{\alpha}(\bm{r}) \sigma^{jk} V_{\alpha}(\bm{r}) \\
    j_{tTo}^i(\bm{r}) &\equiv \sum_{\alpha} \frac{1}{2} \epsilon^{ijk} \tau_3^{(\alpha)}
    \bar{V}_{\alpha}(\bm{r}) \sigma^{jk} V_{\alpha}(\bm{r}),
\end{align}
\end{subequations}
where the subscript $To$ represents the $ij$ components,
that is, the time-odd components of the tensor density.
Note that, since the spin operator is given by
\begin{align}
    \Sigma_i \equiv \gamma_5 \gamma^0 \gamma^i
    = \left(\begin{array}{cc}
        \sigma_i & 0 \\
        0 & \sigma_i
    \end{array}\right),
\end{align}
the third component of the pseudovector current is expressed as
\begin{align}
    j_{PV}^z = \sum_{\alpha}
    V^{\dagger}_{\alpha}(\bm{r}) \Sigma_3 V_{\alpha}(\bm{r}),
\end{align}
which is nothing but the spin density of nucleons.
Therefore, introducing the spin-projection operator
$(1\pm\Sigma_3)/2$ allows us to evaluate the densities
of spin-up and spin-down nucleons,
\begin{align}
\label{eq:def_spin_density}
    \rho_{\uparrow,\downarrow}  = 
    \frac{1}{2} \ev{\psi^{\dagger}(1\pm\Sigma_3)\psi}
    = \frac{1}{2} \left( \rho_B \pm j_{PV}^z \right)
\end{align}

The single-particle Hamiltonian for a system where
time-reversal symmetry is broken is written by
the time-odd densities in Eq. \eqref{eq:densities_odd}
as well as the time-even densities in Eq. \eqref{eq:densities_even} as
\begin{align}
    \hat{h} = & -i\bm{\alpha}\cdot\nabla + \beta M^*(\bm{r})
    + V^0_V(\bm{r}) - \bm{\alpha}\cdot\bm{V}_V(\bm{r}) \notag \\
    & - i \beta \gamma_5 V_{PS}(\bm{r}) - \bm{\Sigma}\cdot\bm{V}_{PV}(\bm{r})
    - i \beta\bm{\alpha} \cdot \bm{V}_{Te}(\bm{r}) \notag \\
    & + \beta\bm{\Sigma}\cdot\bm{V}_{To}(\bm{r}),
\end{align}
where the time-odd potentials are given by
\begin{subequations}
\label{eq:T-odd_potential}
\begin{align}
    \bm{V}_V(\bm{r}) &= \alpha_V \bm{j}_V(\bm{r}) + \tau_3 \alpha_{tV} \bm{j}_{tV}(\bm{r})
    + e \frac{1-\tau_3}{2} \bm{A}(\bm{r}) \\
    V_{PS}(\bm{r}) &= \alpha_{PS} \rho_{PS}(\bm{r}) + \tau_3 \alpha_{tPS} \rho_{tPS}(\bm{r}) \\
    V_{PV}(\bm{r}) &= \alpha_{PV} \bm{j}_{PV}(\bm{r}) + \tau_3 \alpha_{tPV} \bm{j}_{tPV}(\bm{r}) \\
    V_{To}(\bm{r}) &= 2(\alpha_{T} \bm{j}_{To}(\bm{r}) + \tau_3 \alpha_{tT} \bm{j}_{tTo}(\bm{r})).
\end{align}
\end{subequations}

%**************************************************
%        SECTION 3
%**************************************************

\section{The $12^+$ isomeric state of \ce{^{52}F\lowercase{e}} nucleus}
\label{sec:52Fe12+}

Let us next describe the configuration-fixed self-consistent calculation,
which corresponds to the non-collective rotation in the cranking calculation~\cite{szymański1983fast},
for the $12^+$ isomeric state of \ce{^{52}Fe}.
Shell model calculations indicate that
the dominant configuration of the $12^+$ isomeric state
is a simple one in which the angular momenta of valence nucleons in the 1$f_{7/2}$ orbitals 
are coupled to the maximum value
\cite{Ur1998,Gadea2005}.
There is a slight deviation from this simple picture, but such effect can be to some extent 
accounted for in terms of quadrupole deformation. 
Thus, we solve the deformed relativistic Hartree-Fock equation
by fixing the configuration of the valence nucleons. 
For this, we assume that the 6 valence neutrons and the 6 valence protons occupy 
the 1$f_{7/2}$-like single-particle orbitals 
with 
the $m=-7/2$ and $m=-5/2$ orbitals being unoccupied, where $m$ is the $z$ component of the single-particle 
angular momentum. 
To this end, we assume axial symmetry and for simplicity neglect the pairing correlation, 
assuming that Cooper 
pairs in the $^1S_0$ channel are broken
due to the strong alignment of the angular momenta.

In axial symmetric cases,
the single-particle wave function is labeled 
by parity $P$ and the $z$-component of
the angular momentum $j_z$. Thus 
it can be expressed in the cylindrical coordinates $(r_{\perp},\phi,z)$
as
\begin{align}
    \psi_{n,j_z,P}(r_{\perp},\phi,z) = \frac{1}{\sqrt{2\pi}}
    \left( \begin{array}{cc}
        f_{n,j_z,P}^+(r_{\perp},z) e^{i\Lambda_-\phi} \\
        f_{n,j_z,P}^-(r_{\perp},z) e^{i\Lambda_+\phi} \\
        ig_{n,j_z,P}^+(r_{\perp},z) e^{i\Lambda_-\phi} \\
        ig_{n,j_z,P}^-(r_{\perp},z) e^{i\Lambda_+\phi}
    \end{array} \right),
\end{align}
where the index $n$ represents the degrees of freedom other than $j_z$ and $P$
of the single particle states.
Hereafter, we denote the full set of indices for the single particle states
as $\alpha$, that is, $\alpha=\{n,j_z,P\}$.
Then, one can write down the densities in Eq. \eqref{eq:densities_even} as
\begin{subequations}
\begin{align}
    \rho_S(r_{\perp},z) =& \frac{1}{2\pi} \sum_{\alpha} \left(
    {f_{\alpha}^+}^2 + {f_{\alpha}^-}^2 
    - {g_{\alpha}^+}^2 - {g_{\alpha}^-}^2 \right) \\
    \rho_B(r_{\perp},z) =& \frac{1}{2\pi} \sum_{\alpha} \left(
    {f_{\alpha}^+}^2 + {f_{\alpha}^-}^2 
    + {g_{\alpha}^+}^2 + {g_{\alpha}^-}^2 \right) \\
    \bm{j}_{Te} (r_{\perp},z)=& - \frac{1}{\pi} \sum_{\alpha} \left(
    f_{\alpha}^+g_{\alpha}^- + f_{\alpha}^-g_{\alpha}^+
    \right) \bm{e}_{r_\perp} \notag \\
    & - \frac{1}{\pi} \sum_{\alpha} \left(
    f_{\alpha}^+g_{\alpha}^+ - f_{\alpha}^-g_{\alpha}^-
    \right) \bm{e}_z.
\end{align}
\end{subequations}
On the other hand, the densities in Eq. \eqref{eq:densities_odd}
are written as
\begin{align}
    \bm{j}_V(r_{\perp},z)=& \frac{1}{\pi} \sum_{\alpha} \left(
    f_{\alpha}^+g_{\alpha}^- - f_{\alpha}^-g_{\alpha}^+
    \right) \bm{e}_{\phi} \\
    \rho_{PS}(r_{\perp},z) =& - \frac{1}{\pi} \sum_{\alpha} \left(
    f_{\alpha}^+g_{\alpha}^+ + f_{\alpha}^-g_{\alpha}^-
    \right) \\
    \bm{j}_{PV}(r_{\perp},z) =&\frac{1}{\pi} \sum_{\alpha} \left(
    f_{\alpha}^+f_{\alpha}^- + g_{\alpha}^+g_{\alpha}^-
    \right) \bm{e}_{r_\perp} \notag \\
    +& \frac{1}{2\pi} \sum_{\alpha} \left(
    {f_{\alpha}^+}^2 - {f_{\alpha}^-}^2 
    + {g_{\alpha}^+}^2 - {g_{\alpha}^-}^2 \right) \bm{e}_z \\
    \bm{j}_{To}(r_{\perp},z) =& \frac{1}{\pi} \sum_{\alpha} \left(
    f_{\alpha}^+f_{\alpha}^- - g_{\alpha}^+g_{\alpha}^-
    \right) \bm{e}_{r_\perp} \notag \\
    +& \frac{1}{2\pi} \sum_{\alpha} \left(
    {f_{\alpha}^+}^2 - {f_{\alpha}^-}^2 
    - {g_{\alpha}^+}^2 + {g_{\alpha}^-}^2 \right) \bm{e}_z.
\end{align}
Here, the summations are carried out for states below the Fermi energy.
Therefore, the Dirac equation is expressed as
\begin{widetext}
\begin{align}
\label{eq:Dirac_ax}
    &\left( \begin{array}{cccc}
        A^+ & P^+ & 
        \partial_z+D^+_z & 
        \partial_{r_{\perp}}+\Lambda_+/r_{\perp}+D^-_{\perp} \\
        P^+ & A^- &
        \partial_{r_{\perp}}-\Lambda_-/r_{\perp}+D^+_{\perp} & 
        \partial_z+D^-_z \\
        -\partial_z+D^+_z & 
        -\partial_{r_{\perp}}-\Lambda_+/r_{\perp}+D^+_{\perp} & 
        C^+ & P^- \\
        -\partial_{r_{\perp}}+\Lambda_+/r_{\perp}+D^-_{\perp} & 
        -\partial_z+D^-_z & 
        P^- & C^-
    \end{array} \right)
    \left( \begin{array}{c}
    f^+_{\alpha} \\
    f^-_{\alpha} \\
    g^+_{\alpha} \\
    g^-_{\alpha}
    \end{array} \right)
    &= E_{\alpha} \left( \begin{array}{c}
    f^+_{\alpha} \\
    f^-_{\alpha} \\
    g^+_{\alpha} \\
    g^-_{\alpha}
    \end{array} \right),
\end{align}
\end{widetext}
where
\begin{subequations}
\begin{align}
    A^{\pm} &= M^* + V_V^0 \mp V_{PV}^z \pm V_{To}^z \\
    C^{\pm} &= -M^* + V_V^0 \mp V_{PV}^z \mp V_{To}^z \\
    D^{\pm}_{\perp} &= \pm V_V^{\perp} + V_{Te}^{\perp} \\
    D^{\pm}_{z} &= V_{PS} \pm V_{Te}^z \\
    P^{\pm} &= -V_{PV}^{\perp} \pm V_{To}^{\perp}.
\end{align}
\end{subequations}
Note that the subscript '$\perp$' represents
the component perpendicular to the $z$-axis.

Equation \eqref{eq:Dirac_ax} is solved self-consistently in a fixed configuration
to obtain the $12^+$ state.
To this end, 
we fix the configuration with the unoccupied $m=-7/2$ and $m=-5/2$ orbitals 
so that the total $m$ achieves the maximum value of 6 
for both neutrons and protons.
One may also consider an alternative configuration 
in which the $m=+7/2$ and the $m=+5/2$ orbitals are unoccupied.
However, since this configuration is the time-reversal partner
of the former configuration,
only the directions of the total angular momenta differ
while other quantities, such as the binding energies and
the radii, remain the same. 
Notice that the Kramers degeneracy of $\pm j_z$ orbitals is lifted
due to the breaking of the time-reversal symmetry,
that is, the mean-field Dirac equation \eqref{eq:Dirac_ax} depends on
whether the value of $j_z$ is positive or negative.

\begin{figure}[t]
    \centering
    \includegraphics[scale=0.2]{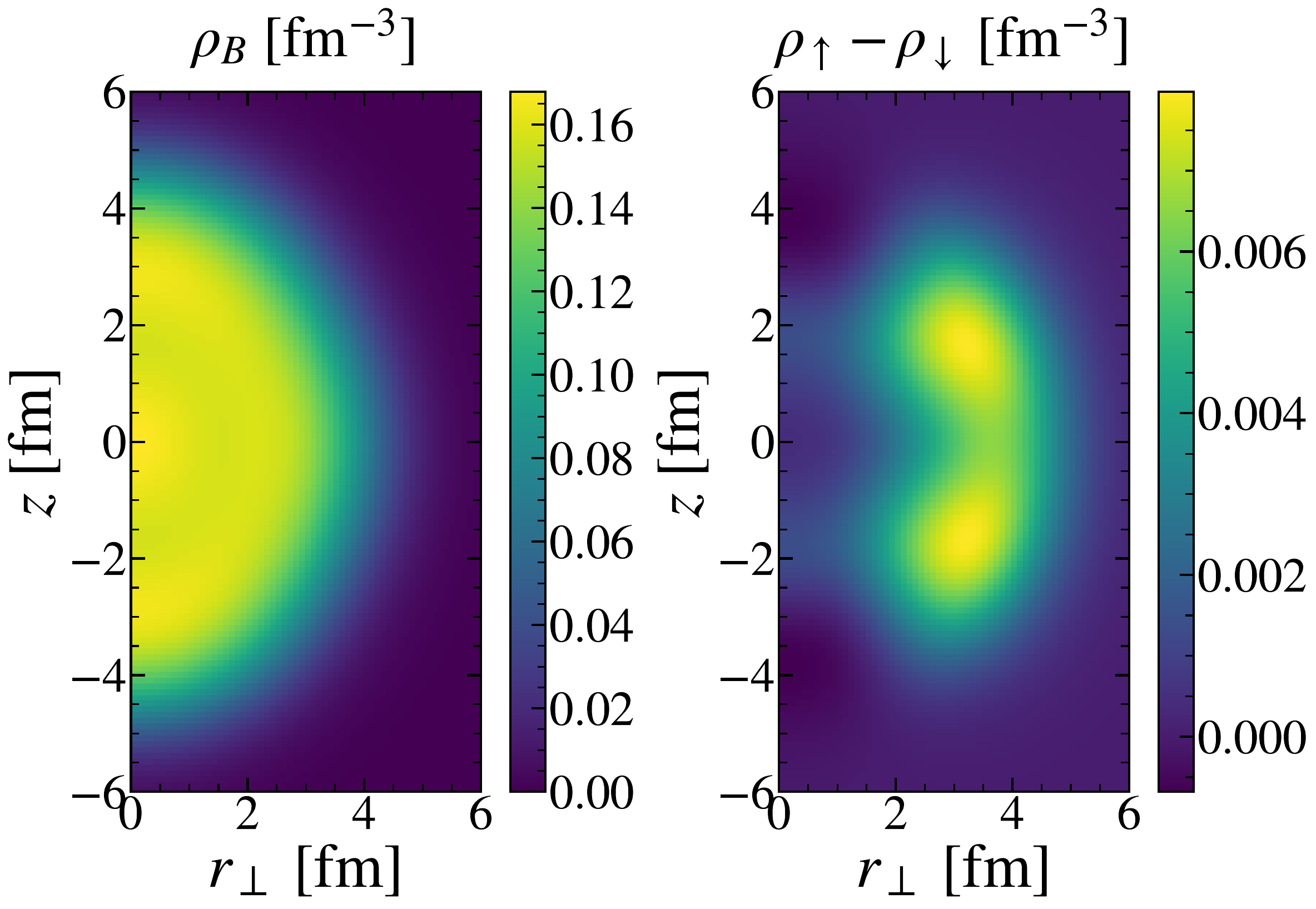}
    \caption{
    Density distribution of the $12^+$ isomeric state 
    of \ce{^{52}Fe} nucleus in the cylindrical coordinate $(r_{\perp},\phi,z)$.
    The left figure represents the nucleon density, 
    while the right figure represents the spin density,
    which is defined as the difference between 
    the density of spin-up nucleons $\rho_{\uparrow}$
    and that of spin-down nucleons $\rho_{\downarrow}$.}
    \label{fig:Fe52}
\end{figure}

Figure \ref{fig:Fe52} shows the calculation results
of the density distribution of the $12^+$ isomeric state of \ce{^{52}Fe} nucleus.
The left figure shows the nucleon density distribution, 
while the right figure shows the spin density distribution, 
which is defined as the difference between 
the density of the spin-up nucleons $\rho_{\uparrow}$
and that of the spin-down nucleons $\rho_{\downarrow}$.
One finds that the spin density is localized
near the surface of the nucleus. 
The radius, the binding energy and the quadrupole deformation
of the $12^+$ isomeric state are summarized in Table \ref{tab:Fe_0+_12+}. 
The table also shows those of the $0^+$ ground state,
which are obtained by the axial symmetric HFB calculation
also with the PCF-PK1 parameter set.
The quadrupole deformation parameter $\beta_2$ in the table is given by
\begin{align}
    \beta_2 = \frac{\sqrt{5\pi}}{3AR_0^2} Q_{20},
\end{align}
where $R_0 = 1.2 A^{1/3}$~fm is a typical nuclear radius
for mass number $A$.
Here, the quadrupole moment $Q_{20}$ is defined as
\begin{align}
    Q_{20} = \sqrt{\frac{16\pi}{5}} \int d^3\bm{r} r^2Y_{20}(\Omega)\rho_B(\bm{r}).
\end{align}
Table \ref{tab:Fe_0+_12+} indicates that
the quadrupole deformation of the $12^+$ state is suppressed from 0.255 to 0.164 
compared to the $0^+$ ground state.
From the calculated binding energies, 
we obtain the excitation energy of the $12^+$ isomeric state
to be $6.18$ MeV without any additional fitting parameter,
which is close to the experimental value
of $6.96$ MeV.

\begin{figure}[t]
    \centering
    \includegraphics[scale=0.2]{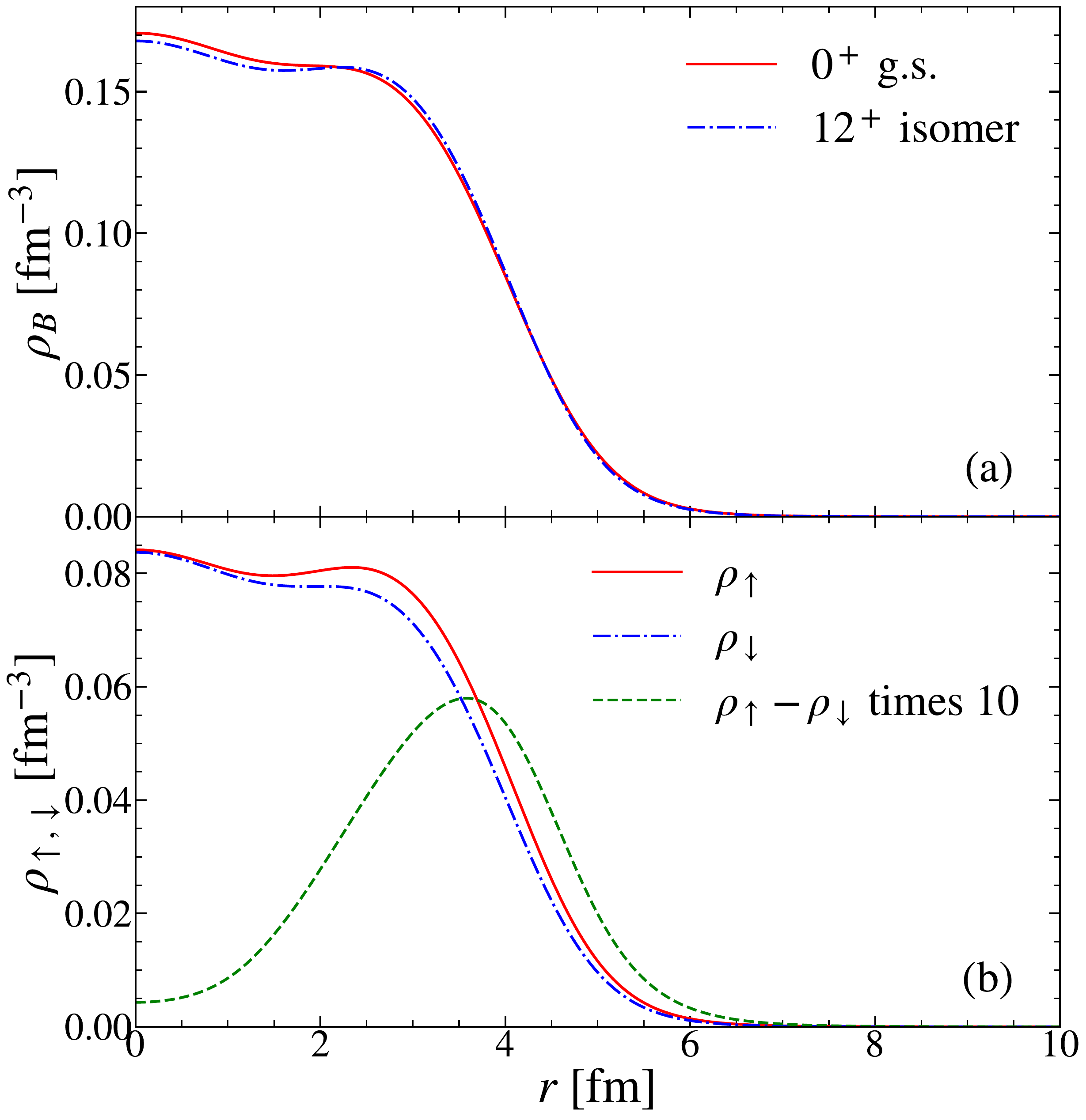}
    \caption{The angular average of the nucleon and the spin density distributions 
    of the $12^+$ isomeric state of the \ce{^{52}Fe} nucleus. 
    Those of the $0^+$ ground state are also plotted for comparison.
    In the upper figure (a), 
    the red solid line represents the $0^+$ ground state
    while the blue dashed line represents the $12^+$ isomeric state.
    In the lower figure (b),
    the red solid (the blue dashed) line represents the density distribution
    of the spin-up (the spin-down) nucleons. 
    The green dotted line represents the spin density defined by
    the difference between the density of spin-up nucleons
    and that of spin-down nucleons.  }
    \label{fig:ang_av}
\end{figure}

Figures \ref{fig:ang_av} shows the angular average of 
the density distribution of $12^+$ isomeric state
as well as that of the $0^+$ ground state for comparison.
The angular average of the density distribution
$\rho(r_{\perp},z)$ is defined as the angular integration,
\begin{align}
    \rho_{\text{av}}(r) = \frac{1}{4\pi} 
    \int d\Omega \,\rho(r_\perp = r \sin{\theta},
    z = r \cos{\theta}),
\end{align}
where we use the spherical coordinate $(r,\theta,\phi)$.
The upper panel shows that the $12^+$ isomeric state is slightly
smaller than the $0^+$ ground state,
which is consistent with their radii listed in Table \ref{tab:Fe_0+_12+}.
The lower panel of Fig. \ref{fig:ang_av}
also indicates that  the spin density is localized near the surface.

Let us discuss what causes the reduction of the deformation for the $12^+$ isomeric state.
An obvious difference between the $0^+$ ground state and the $12^+$ isomeric state
is that the pairing correlation is taken into account in 
the former while it is neglected in the latter.
However, in general, 
the pairing correlation tends to reduce the deformation, and 
therefore it may be difficult to attribute the observed reduction of the deformation parameter 
entirely to the pairing correlation. 
There are in fact three other possible origins for the reduction of the deformation parameter: 
(i) the configuration fixing within the $1f_{7/2}$ orbital,
(ii) the self-consistent response of the mean-field potential to the configuration fixing,
and (iii) the contribution from the time-odd potentials.
In fact, these three factors are not independent,
since there is no field response without the configuration fixing
and the time-odd potentials do not arise
unless one takes into account the self-consistency of the mean-field potential.

To separate these three contributions,
we carry out deformed HF calculations with several different settings. 
The shift of the quadrupole deformations is summarized in Table \ref{tab:deform_shift}.
First, we switch off the pairing correlation in the $0^+$ ground state (see the second 
row in the table).
As expected, this slightly increases the deformation, that is the opposite direction to 
the difference of the deformation parameter between the $0^+$ and the $12^+$ states. 
Second, we excite one neutron and one proton in the $0^+$ ground state
without the pairing interaction.
Specifically, we remove a neutron and a proton from the $j_z = -5/2$ orbitals
in the $1f_{7/2}$-like states and add them to the $j_z = +7/2$ orbitals to form the $12^+$ state. 
As one can see in the third row in the table, this reduces the deformation from $\beta_2=0.261$ to 0.232, since the single-particle wave functions for the $|j_z|=7/2$ orbitals
are less prolate than those for $|j_z|=5/2$ orbitals.
That is nothing but a geometrical effect corresponding to the origin (i).
Third, to isolate the origin (ii), we perform the configuration-fixed self-consistent calculation
neglecting the time-odd potentials in Eq. \eqref{eq:T-odd_potential}.
This amplifies the reduction of the deformation, that is, 
by taking into account the self-consistency the deformation parameter reduces from 0.232 
to 0.174 (see the fourth row in the table). 
Finally, we include the origin (iii), that is, the time-odd potentials.
This calculation yields the results of Fig. \ref{fig:Fe52}
and Table \ref{tab:Fe_0+_12+} with $\beta_2=0.164$. 
According to Table \ref{tab:deform_shift},
one can see that the dominant contributions to the deformation change 
are the geometrical effect and the self-consistency,
while the time-odd potentials provide smaller corrections.

\begin{table}[t]
    \centering
    \caption{
    The radii, the binding energy, the pairing gaps and
    the quadrupole deformations $\beta_2$ for the $0^+$ ground state and for the $12^+$ isomeric state
    of the \ce{^{52}Fe} nucleus.
    }
    \begin{tabular}{c|cc}
    \hline \hline
    & $0^+$ g.s. & $12^+$ isomer \\ \hline
    Radius [fm] & $3.633$ & $3.614$ \\
    Binding Energy [MeV] & $-442.36$ & $-436.18$ \\
    Pairing Gap [MeV] & $1.851$ & $-$ \\ 
    $\beta_2$ & $0.255$ & $0.164$ \\
    \hline \hline
    \end{tabular}
\label{tab:Fe_0+_12+}
\end{table}

\begin{table*}[t]
    \centering
    \setlength{\tabcolsep}{10pt}
    \caption{The quadrupole deformation parameters and their changes,
    obtained with several setups of the calculation as explained in the main text.
    $\Delta \beta_2$ denotes the change with respect to the preceding row.}
    \begin{tabular}{ccccc|cc}
    \hline \hline
    $I^\pi$ & pairing & configuration & self-consistency & time-odd fields & $\beta_2$ & $\Delta \beta_2$ \\ \hline
    $0^+$ & yes & 0p0h & yes & - & $0.255$ & - \\
    $0^+$ & no & 0p0h & yes & - & $0.261$ & $+0.006$ \\
    $12^+$& no & 2p2h & no & no & $0.232$ & $-0.029$ \\ 
    $12^+$& no & 2p2h & yes & no & $0.174$ & $-0.058$ \\
    $12^+$& no & 2p2h & yes & yes & $0.164$ & $-0.010$ \\
    \hline \hline
    \end{tabular}
\label{tab:deform_shift}
\end{table*}

%**************************************************
%        SECTION 4
%**************************************************

\section{correlation between the spin slope parameter
and the spin skin thickness}
\label{sec:correlation}

We now turn to the spin ``skin thickness"
of \ce{^{52}Fe},  
which is expected to be correlated with the spin slope parameter, $L_s$. 
For this purpose,
we consider three possible definitions of the spin ``skin thickness". 
Notice that the neutron skin thickness $\Delta r_{np}$ is defined as
\begin{align}
    \Delta r_{np} = r_n - r_p,
\end{align}
where $r_{n,p}$ refer to the root-mean-square (rms) radii of neutrons and protons,
\begin{align}
    r_{n,p} = \left[\frac{\int d^3\bm{r} r^2 \rho_{n,p}(\bm{r})}
    {\int d^3\bm{r} \rho_{n,p}(\bm{r})} \right]^{\frac{1}{2}}.
\end{align}
As a direct analogy to the neutron skin thickness, 
we first define the spin skin thickness as 
the difference in radii
between the spin-up nucleons and the spin-down nucleons
in the $12^+$ isomeric state of \ce{^{52}Fe},
\begin{align}
\label{eq:def_skin1}
    \Delta r_s = r_{\uparrow} - r_{\downarrow},
\end{align}
where the radii of the spin-up (the spin-down) nucleons are defined as
\begin{align}
    r_{\uparrow,\downarrow} = 
    \left[\frac{\int d^3\bm{r} r^2 \rho_{\uparrow,\downarrow}(\bm{r})}
    {\int d^3\bm{r} \rho_{\uparrow,\downarrow}(\bm{r})} \right]^{\frac{1}{2}}.
\end{align}
Notice that $\Delta r_s$ is determined only by the properties
of the $12^+$ isomeric state.

Since $\Delta r_s$ would be difficult to measure experimentally, 
we also consider  two additional definitions of the spin skin thickness.
One is the difference in the rms radii
between the $0^+$ ground state and the $12^+$ isomeric state,
\begin{align}
\label{eq:def_skin2}
    \Delta r'_s =r(12^+)-r(0^+) .
\end{align}
Strictly speaking, $\Delta r'_s$ is not a skin thickness
but simply the difference in radii between different states.
However, 
if the radius of the $12^+$ isomeric state 
of the \ce{^{52}Fe} nucleus is measured experimentally,
one can easily derive this quantity.
Notice that 
we here define $\Delta r'_s$ through the matter radius,
which is relevant to 
hadron-scattering experiments with isomeric beams.
The other definition is the difference in mean-square charge radii 
between the $0^+$ ground state and the $12^+$ isomeric state,
\begin{align}
\label{eq:def_skin3}
    \Delta r^2_{ch} = r_{ch}^2(12^+) -  r_{ch}^2(0^+).
\end{align}
Here, the charge radius is defined as
\begin{align}
    r_{ch} = \sqrt{r_p^2 + l_p^2},
\end{align}
where $l_p$ is an empirical charge radius of protons given by 
$l_p = 0.8$ fm.
A similar quantity can be defined with the matter radii, 
but here we used the charge radius, 
as it is a direct observable for laser-spectroscopy experiments.

In Ref. \cite{Tachibana2025}, 
we calculated the EOS of spin-polarized symmetric nuclear matter
with the relativistic point-coupling model \cite{PCF-PK1} 
and discussed the spin slope parameter, $L_s$.  
For the isoscalar polarization
where neutrons and protons are spin-polarized
in the same direction, as in the case for the $12^+$ isomeric state of \ce{^{52}Fe}, 
one can introduce the spin polarization rate $\Delta$ defined as
\begin{align}
    \Delta = \frac{\rho_{\uparrow} - \rho_{\downarrow}}{\rho_B}, 
\end{align}
to characterize the EOS.
Notice that 
with the definitions of the densities of spin-up and spin-down nucleons 
given by Eq. \eqref{eq:def_spin_density},
the spin polarization rate $\Delta$ is
given by $j_{PV}^z / \rho_B$.
The EOS can be expanded with respect to 
the spin polarization rate $\Delta$ as,
\begin{align}
    \frac{E}{A}(\rho_B,\delta,\Delta) = \frac{E}{A}(\rho_B,\delta,\Delta=0)
    + W(\rho_B,\delta) \Delta^2 + \mathcal{O}(\Delta^4),
\end{align}
where $\delta$ is defined as $\delta=(\rho_n-\rho_p)/\rho_B$. 
Here, 
the spin symmetry energy $W(\rho_B,\delta)$ is defined as
\begin{align}
    W(\rho_B,\delta) = \frac{1}{2}
    \left. \frac{\partial^2}{\partial\Delta^2}
    \frac{E}{A} \right|_{\Delta=0}.
\end{align}
The spin slope parameter $L_s$ is defined as the derivative 
of the spin symmetry energy,
\begin{align}
    L_s = 3\rho_0 \left. 
    \frac{\partial W}{\partial \rho_B}
    \right|_{\rho_B=\rho_0},
\end{align}
where $\rho_0$ refers to the saturation density.

Our interest here is to discuss the correlation between the spin slope parameter $L_s$ 
and the spin skin thickness. 
For this purpose, 
we apply a similar method to that proposed in Ref. \cite{Inakura2015},
in which the density dependence of the coupling constant is varied.
To this end,
we introduce a new parameter $y$
and modify the coupling constant for the isovector-vector channel
$\alpha_{tV}$ as
\begin{align}
    \alpha_{tV}(\rho_B) \rightarrow (1-y) \alpha_{tV}(\rho_B)
    + y \alpha_{tV}(\rho_0).
    \label{eq:y}
\end{align}
Notice that this modification keeps the value of $\alpha_{tV}$ at the
saturation density $\rho_0$.
It also changes the isoscalar time-odd potential in nuclear matter
and high-spin isomers through the relation among the coupling constants
\eqref{eq:rel_cc},
and it therefore changes the spin slope parameter \cite{Tachibana2025}.

\begin{figure}[t]
    \centering
    \includegraphics[scale=0.2]{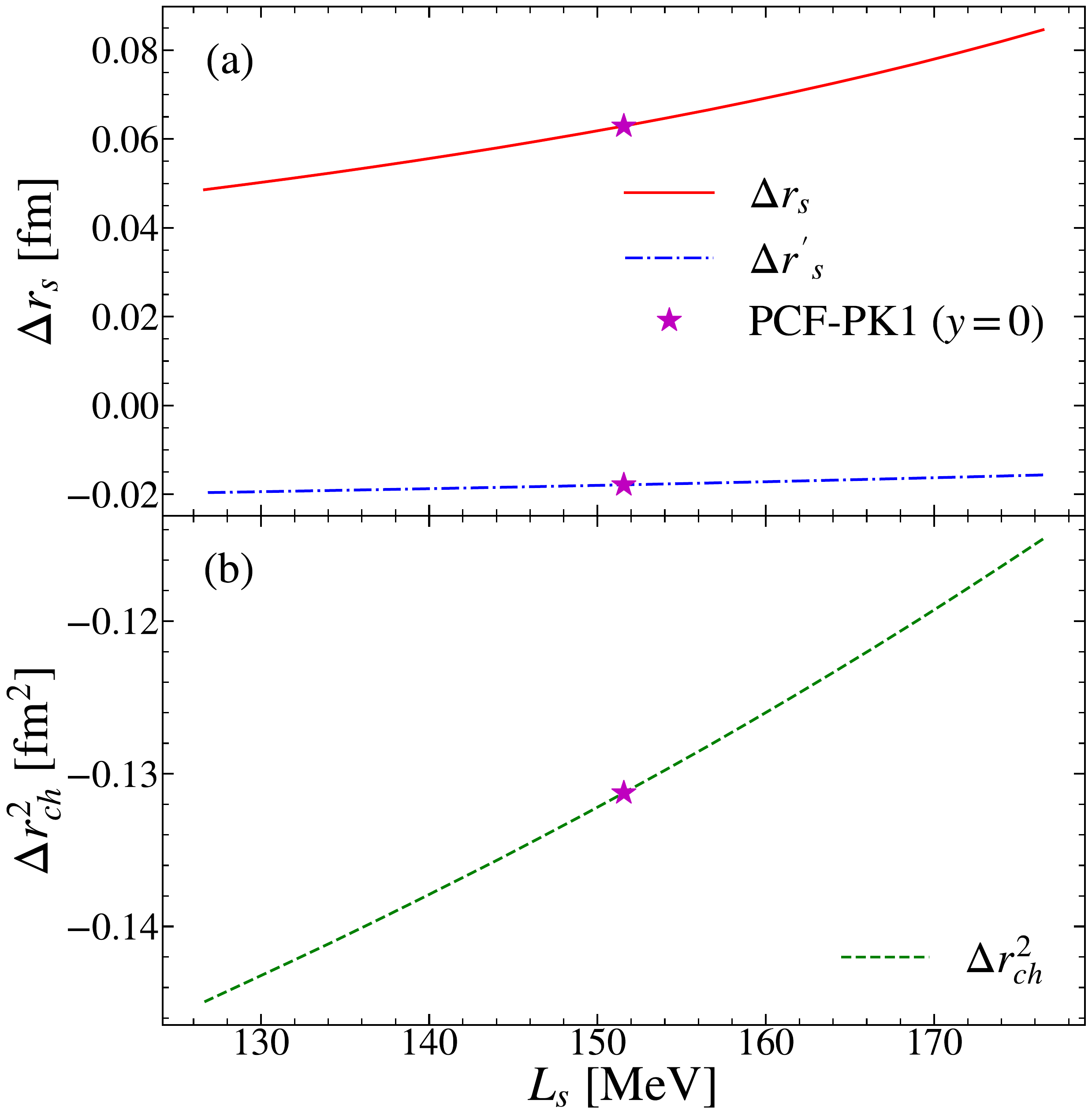}
    \caption{The correlation between the spin slope parameter
    for isoscalar polarization and the spin skin thickness of \ce{^{52}Fe}.
    In the upper panel (a), the red solid line represents $\Delta r_s$ 
    defined by Eq. \eqref{eq:def_skin1},
    while the blue dashed-dotted line represents $\Delta r'_s$
    defined by Eq. \eqref{eq:def_skin2}.
    In the lower panel (b),
    the green dashed line represents $\Delta r^2_{ch}$
    defined by Eq. \eqref{eq:def_skin3}.
    The stars show the result with the original PCF-PK1 parameter set,
    that is, with $y = 0$ in Eq. \eqref{eq:y}.}
    \label{fig:spin_skin}
\end{figure}

Figure \ref{fig:spin_skin} shows the numerical results of
the correlation between the spin slope parameter
and the spin skin thickness of \ce{^{52}Fe}.
In the upper panel (a), the red solid line represents the spin skin thickness $\Delta r_s$
defined by Eq. \eqref{eq:def_skin1},
while the blue dashed-dotted line represents the spin skin thickness $\Delta r'_s$
defined by Eq. \eqref{eq:def_skin2}.
On the other hand, in the lower panel (b),
the green dashed line represents the spin skin thickness $\Delta r^2_{ch}$
defined by Eq. \eqref{eq:def_skin3}.
One can see that all the definitions of the spin skin thickness
exhibit positive correlations with the spin slope parameter, 
even though the sensitivity to $L_s$ are largely different among the three definitions.
The size of $\Delta r_s$ and its dependence on the spin slope parameter
are larger than those of $\Delta r'_s$,
since $\Delta r_s$ reflects the spin profile in the $12^+$ isomeric state
but this is not the case for $\Delta r'_s$.
Although the $\Delta r_s$ would be difficult to observe
in scattering experiments,
$\Delta r_s$ has a strong correlation
with the spin slope parameter 
and therefore can be a better probe
for the spin slope parameter.
Based on the liner fitting,
an accuracy of $\pm0.02$ fm for $\Delta r_s$ is required
to constrain the spin slope parameter $L_s$ within the range of $\pm30$ MeV,
which is comparable to the current uncertainty of the slope parameter $L$ 
\cite{Oertel2017}.
This accuracy has the same order as the experimental error
for the neutron skin thickness of the \ce{^{208}Pb} nucleus
deduced from the proton-elastic-scattering experiment \cite{Zenihiro}.

In contrast,
a higher accuracy of about $\pm0.002$ fm is required for $\Delta r'_s$
to determine the spin slope parameter $L_s$ within the same range.
This weaker correlation between $\Delta r'_s$
and the spin slope parameter $L_s$ originates from
the fact that the time-odd interactions do not have a large impact 
on the spin-averaged properties of the $12^+$ isomer,
such as the radius and the deformation.
For instance, assuming a sharp-cut and volume-conserving density distribution,
the rms radius for a deformed nucleus as a function of $\beta_2$ is given by
\begin{align}
\label{eq:rad_of_beta}
    R(\beta_2) = \sqrt{\frac{3}{5}}R_0 \left[
    1 + \frac{5}{8\pi}\beta_2^2 + \frac{25\sqrt{5}}{168\pi^{3/2}}\beta_2^3
    +\mathcal{O}(\beta_2^4)\right].
\end{align}
From Eq. \eqref{eq:rad_of_beta} and Table \ref{tab:Fe_0+_12+},
one obtains the relative difference in the rms radii as $[R(\beta_2(12^+))-R(\beta_2(0^+))]/R(\beta_2(0^+))\simeq-0.8\%$.
Since the actual value is $[r_{\text{rms}}(12^+)-r_{\text{rms}}(0^+)]/r_{\text{rms}}(0^+)\simeq-0.5\%$,
one can find that a major factor for the contraction of the $12^+$ isomer 
is the reduction of the quadrupole deformation.
Since, as discussed in Sec. \ref{sec:52Fe12+},
the contributions of the time-odd potentials to the deformation
are smaller than the other factors,
the dependence of $\Delta r'_s$ on $L_s$ is weak.

We mention that
$\Delta r^2_{ch}$ has the same physical meaning as $\Delta r'_s$,
and therefore its correlation with the spin slope parameter $L_s$
is as weak as that of $\Delta r'_s$.
However,
by taking square of the radii, the absolute value of $\Delta r^2_{ch}$ is much larger than 
$\Delta r'_s$, and so is their dependence on $L_s$. 
Given
a high resolution of the laser-spectroscopy experiments, 
it is expected that $\Delta r^2_{ch}$ can serve as a good probe of $L_s$.
Notice that the experimental data for the differences in mean-square charge radii 
between the ground states
and the high-spin isomeric states have errors of $0.01$ fm$^2$ for the \ce{^{42}Sc} \cite{Koszorus2021}
and of $0.002$ fm$^2$ of the \ce{^{129}In} \cite{Vernon2025}.
If a similar accuracy is achieved for the \ce{^{52}Fe} nucleus,
the spin slope parameter is determined with the uncertainty of $\pm (3-15)$ MeV.
This range is narrower than that of the slope parameter $L$.

%**************************************************
%        SECTION 5
%**************************************************

\section{SUMMARY}
\label{sec:summary}

We have explored a possible experimental probe
for constraining the EOS of spin-polarized nuclear matter.
In this study,
we focused on the spin slope parameter discussed in the Ref. \cite{Tachibana2025}
and considered a quantity correlated with the spin slope parameter.
To this end,
we have calculated the $0^+$ ground state
and the $12^+$ isomeric state of the \ce{^{52}Fe} nucleus.
As a candidate of experimental probes
for the spin slope parameter,
we considered three possible definitions of the spin skin thickness
of the \ce{^{52}Fe} nucleus.
The first definition is the difference in rms radii
between the spin-up and the spin-down nucleons
in the $12^+$ isomeric state.
The second definition is 
the difference in rms radii
between the $0^+$ ground state and the $12^+$ isomeric state
of the \ce{^{52}Fe} nucleus.
The third definition is 
the difference in the mean-square charge radii
between the $0^+$ ground state and the $12^+$ isomeric state.
We found that 
all of these definitions exhibit positive correlations with
the spin slope parameter.
The first definition has a strong correlation with
the spin slope parameter and may serve as a good probe,
although it remains challenging
to detect experimentally the spin skin thickness in this definition.
On the other hand,
the second definition of the spin skin thickness and their dependence on
the spin slope parameter are much smaller than those of the first definition.
Even though the third definition carries the same information as the second 
definition, by taking the square both the absolute value and the dependence increase. 
Given the high resolution of the laser-spectroscopy experiments, 
it would therefore 
serve as a good probe of $L_s$.  

In this study,
we have neglected the pairing correlation 
in the $12^+$ isomeric state of the \ce{^{52}Fe} nucleus.
Although blocking certain orbitals may hinder the pairing effects,
we should clarify the role of the pairing correlation 
in high-spin isomer.
Also, it is an experimental challenge to observe 
the spin skin thickness $\Delta r_s$
defined by Eq. \eqref{eq:def_skin1}  
in scattering and collision experiments. 
In this connection, it might be interesting to investigate the correlation 
between the Gamow-Teller giant resonances from the $12^+$ state discussed 
in Ref. \cite{Liang2018} and $\Delta r_s$. 
Furthermore,
it would be also important to propose a probe
for constraining the $W(\rho_0,\delta)$ itself, 
that is, the spin-symmetry energy at the saturation density. 
These are left for interesting future works.

%**************************************************
%        ACKNOWLEDFEMENT
%**************************************************

\section*{acknowledgments}
We thank M. Cheoun, G. Col\`o, N. Hinohara, S. Mizutori, J. Zenihiro, M. Dozono, H. Togashi, K. Uzawa, X. Roca-Maza and P.W. Zhao for useful discussions. 
This work was supported by the JSPS KAKENHI (Grant Nos.
JP23K03414, JP23H05434, JP25H01269, JP25K07322, 25K24695, and 26H01416),
JST ERATO (Grant No. JPMJER2304),
JST SPRING (Grant No. JPMJSP2110),
the funding of the China Institute of Atomic Energy (Grants No. YC010270525794 and No. PA010271225779).

\bibliographystyle{apsrev4-2.bst}
\bibliography{main}% Produces the bibliography via BibTeX.

\providecommand{\noopsort}[1]{}\providecommand{\singleletter}[1]{#1}%
\begin{thebibliography}{38}%
\makeatletter
\providecommand \@ifxundefined [1]{%
 \@ifx{#1\undefined}
}%
\providecommand \@ifnum [1]{%
 \ifnum #1\expandafter \@firstoftwo
 \else \expandafter \@secondoftwo
 \fi
}%
\providecommand \@ifx [1]{%
 \ifx #1\expandafter \@firstoftwo
 \else \expandafter \@secondoftwo
 \fi
}%
\providecommand \natexlab [1]{#1}%
\providecommand \enquote  [1]{``#1''}%
\providecommand \bibnamefont  [1]{#1}%
\providecommand \bibfnamefont [1]{#1}%
\providecommand \citenamefont [1]{#1}%
\providecommand \href@noop [0]{\@secondoftwo}%
\providecommand \href [0]{\begingroup \@sanitize@url \@href}%
\providecommand \@href[1]{\@@startlink{#1}\@@href}%
\providecommand \@@href[1]{\endgroup#1\@@endlink}%
\providecommand \@sanitize@url [0]{\catcode `\\12\catcode `\$12\catcode `\&12\catcode `\#12\catcode `\^12\catcode `\_12\catcode `\%12\relax}%
\providecommand \@@startlink[1]{}%
\providecommand \@@endlink[0]{}%
\providecommand \url  [0]{\begingroup\@sanitize@url \@url }%
\providecommand \@url [1]{\endgroup\@href {#1}{\urlprefix }}%
\providecommand \urlprefix  [0]{URL }%
\providecommand \Eprint [0]{\href }%
\providecommand \doibase [0]{https://doi.org/}%
\providecommand \selectlanguage [0]{\@gobble}%
\providecommand \bibinfo  [0]{\@secondoftwo}%
\providecommand \bibfield  [0]{\@secondoftwo}%
\providecommand \translation [1]{[#1]}%
\providecommand \BibitemOpen [0]{}%
\providecommand \bibitemStop [0]{}%
\providecommand \bibitemNoStop [0]{.\EOS\space}%
\providecommand \EOS [0]{\spacefactor3000\relax}%
\providecommand \BibitemShut  [1]{\csname bibitem#1\endcsname}%
\let\auto@bib@innerbib\@empty
%</preamble>
\bibitem [{\citenamefont {Yamamoto}\ \emph {et~al.}(2017)\citenamefont {Yamamoto}, \citenamefont {Togashi}, \citenamefont {Tamagawa}, \citenamefont {Furumoto}, \citenamefont {Yasutake},\ and\ \citenamefont {Rijken}}]{Yamamoto_Togashi}%
  \BibitemOpen
  \bibfield  {author} {\bibinfo {author} {\bibfnamefont {Y.}~\bibnamefont {Yamamoto}}, \bibinfo {author} {\bibfnamefont {H.}~\bibnamefont {Togashi}}, \bibinfo {author} {\bibfnamefont {T.}~\bibnamefont {Tamagawa}}, \bibinfo {author} {\bibfnamefont {T.}~\bibnamefont {Furumoto}}, \bibinfo {author} {\bibfnamefont {N.}~\bibnamefont {Yasutake}},\ and\ \bibinfo {author} {\bibfnamefont {T.~A.}\ \bibnamefont {Rijken}},\ }\href {https://doi.org/10.1103/PhysRevC.96.065804} {\bibfield  {journal} {\bibinfo  {journal} {Phys. Rev. C}\ }\textbf {\bibinfo {volume} {96}},\ \bibinfo {pages} {065804} (\bibinfo {year} {2017})}\BibitemShut {NoStop}%
\bibitem [{\citenamefont {Togashi}\ \emph {et~al.}(2017)\citenamefont {Togashi}, \citenamefont {Nakazato}, \citenamefont {Takehara}, \citenamefont {Yamamuro}, \citenamefont {Suzuki},\ and\ \citenamefont {Takano}}]{Togashi_2017}%
  \BibitemOpen
  \bibfield  {author} {\bibinfo {author} {\bibfnamefont {H.}~\bibnamefont {Togashi}}, \bibinfo {author} {\bibfnamefont {K.}~\bibnamefont {Nakazato}}, \bibinfo {author} {\bibfnamefont {Y.}~\bibnamefont {Takehara}}, \bibinfo {author} {\bibfnamefont {S.}~\bibnamefont {Yamamuro}}, \bibinfo {author} {\bibfnamefont {H.}~\bibnamefont {Suzuki}},\ and\ \bibinfo {author} {\bibfnamefont {M.}~\bibnamefont {Takano}},\ }\href {https://doi.org/https://doi.org/10.1016/j.nuclphysa.2017.02.010} {\bibfield  {journal} {\bibinfo  {journal} {Nucl. Phys. A}\ }\textbf {\bibinfo {volume} {961}},\ \bibinfo {pages} {78} (\bibinfo {year} {2017})}\BibitemShut {NoStop}%
\bibitem [{\citenamefont {Shen}\ \emph {et~al.}(1998)\citenamefont {Shen}, \citenamefont {Toki}, \citenamefont {Oyamatsu},\ and\ \citenamefont {Sumiyoshi}}]{Shen}%
  \BibitemOpen
  \bibfield  {author} {\bibinfo {author} {\bibfnamefont {H.}~\bibnamefont {Shen}}, \bibinfo {author} {\bibfnamefont {H.}~\bibnamefont {Toki}}, \bibinfo {author} {\bibfnamefont {K.}~\bibnamefont {Oyamatsu}},\ and\ \bibinfo {author} {\bibfnamefont {K.}~\bibnamefont {Sumiyoshi}},\ }\href {https://doi.org/https://doi.org/10.1016/S0375-9474(98)00236-X} {\bibfield  {journal} {\bibinfo  {journal} {Nucl. Phys. A}\ }\textbf {\bibinfo {volume} {637}},\ \bibinfo {pages} {435} (\bibinfo {year} {1998})}\BibitemShut {NoStop}%
\bibitem [{\citenamefont {Hempel}\ \emph {et~al.}(2012)\citenamefont {Hempel}, \citenamefont {Fischer}, \citenamefont {Schaffner-Bielich},\ and\ \citenamefont {Liebendörfer}}]{Hempel_2012}%
  \BibitemOpen
  \bibfield  {author} {\bibinfo {author} {\bibfnamefont {M.}~\bibnamefont {Hempel}}, \bibinfo {author} {\bibfnamefont {T.}~\bibnamefont {Fischer}}, \bibinfo {author} {\bibfnamefont {J.}~\bibnamefont {Schaffner-Bielich}},\ and\ \bibinfo {author} {\bibfnamefont {M.}~\bibnamefont {Liebendörfer}},\ }\href {https://doi.org/10.1088/0004-637X/748/1/70} {\bibfield  {journal} {\bibinfo  {journal} {Astrophys. J.}\ }\textbf {\bibinfo {volume} {748}},\ \bibinfo {pages} {70} (\bibinfo {year} {2012})}\BibitemShut {NoStop}%
\bibitem [{\citenamefont {Sumiyoshi}\ \emph {et~al.}(2005)\citenamefont {Sumiyoshi}, \citenamefont {Yamada}, \citenamefont {Suzuki}, \citenamefont {Shen}, \citenamefont {Chiba},\ and\ \citenamefont {Toki}}]{Sumiyoshi_2005}%
  \BibitemOpen
  \bibfield  {author} {\bibinfo {author} {\bibfnamefont {K.}~\bibnamefont {Sumiyoshi}}, \bibinfo {author} {\bibfnamefont {S.}~\bibnamefont {Yamada}}, \bibinfo {author} {\bibfnamefont {H.}~\bibnamefont {Suzuki}}, \bibinfo {author} {\bibfnamefont {H.}~\bibnamefont {Shen}}, \bibinfo {author} {\bibfnamefont {S.}~\bibnamefont {Chiba}},\ and\ \bibinfo {author} {\bibfnamefont {H.}~\bibnamefont {Toki}},\ }\href {https://doi.org/10.1086/431788} {\bibfield  {journal} {\bibinfo  {journal} {Astrophys. J.}\ }\textbf {\bibinfo {volume} {629}},\ \bibinfo {pages} {922} (\bibinfo {year} {2005})}\BibitemShut {NoStop}%
\bibitem [{\citenamefont {Oertel}\ \emph {et~al.}(2017)\citenamefont {Oertel}, \citenamefont {Hempel}, \citenamefont {Kl\"ahn},\ and\ \citenamefont {Typel}}]{Oertel2017}%
  \BibitemOpen
  \bibfield  {author} {\bibinfo {author} {\bibfnamefont {M.}~\bibnamefont {Oertel}}, \bibinfo {author} {\bibfnamefont {M.}~\bibnamefont {Hempel}}, \bibinfo {author} {\bibfnamefont {T.}~\bibnamefont {Kl\"ahn}},\ and\ \bibinfo {author} {\bibfnamefont {S.}~\bibnamefont {Typel}},\ }\href {https://doi.org/10.1103/RevModPhys.89.015007} {\bibfield  {journal} {\bibinfo  {journal} {Rev. Mod. Phys.}\ }\textbf {\bibinfo {volume} {89}},\ \bibinfo {pages} {015007} (\bibinfo {year} {2017})}\BibitemShut {NoStop}%
\bibitem [{\citenamefont {Akmal}\ \emph {et~al.}(1998)\citenamefont {Akmal}, \citenamefont {Pandharipande},\ and\ \citenamefont {Ravenhall}}]{APR}%
  \BibitemOpen
  \bibfield  {author} {\bibinfo {author} {\bibfnamefont {A.}~\bibnamefont {Akmal}}, \bibinfo {author} {\bibfnamefont {V.~R.}\ \bibnamefont {Pandharipande}},\ and\ \bibinfo {author} {\bibfnamefont {D.~G.}\ \bibnamefont {Ravenhall}},\ }\href {https://doi.org/10.1103/PhysRevC.58.1804} {\bibfield  {journal} {\bibinfo  {journal} {Phys. Rev. C}\ }\textbf {\bibinfo {volume} {58}},\ \bibinfo {pages} {1804} (\bibinfo {year} {1998})}\BibitemShut {NoStop}%
\bibitem [{\citenamefont {Togashi}\ and\ \citenamefont {Takano}(2013)}]{Togashi_2013}%
  \BibitemOpen
  \bibfield  {author} {\bibinfo {author} {\bibfnamefont {H.}~\bibnamefont {Togashi}}\ and\ \bibinfo {author} {\bibfnamefont {M.}~\bibnamefont {Takano}},\ }\href {https://doi.org/https://doi.org/10.1016/j.nuclphysa.2013.02.014} {\bibfield  {journal} {\bibinfo  {journal} {Nucl. Phys. A}\ }\textbf {\bibinfo {volume} {902}},\ \bibinfo {pages} {53} (\bibinfo {year} {2013})}\BibitemShut {NoStop}%
\bibitem [{\citenamefont {Bombaci}\ and\ \citenamefont {Lombardo}(1991)}]{Bompaci_1991}%
  \BibitemOpen
  \bibfield  {author} {\bibinfo {author} {\bibfnamefont {I.}~\bibnamefont {Bombaci}}\ and\ \bibinfo {author} {\bibfnamefont {U.}~\bibnamefont {Lombardo}},\ }\href {https://doi.org/10.1103/PhysRevC.44.1892} {\bibfield  {journal} {\bibinfo  {journal} {Phys. Rev. C}\ }\textbf {\bibinfo {volume} {44}},\ \bibinfo {pages} {1892} (\bibinfo {year} {1991})}\BibitemShut {NoStop}%
\bibitem [{\citenamefont {Lattimer}\ and\ \citenamefont {Lim}(2013)}]{Lattimer_2013}%
  \BibitemOpen
  \bibfield  {author} {\bibinfo {author} {\bibfnamefont {J.~M.}\ \bibnamefont {Lattimer}}\ and\ \bibinfo {author} {\bibfnamefont {Y.}~\bibnamefont {Lim}},\ }\href {https://doi.org/10.1088/0004-637X/771/1/51} {\bibfield  {journal} {\bibinfo  {journal} {Astrophys. J.}\ }\textbf {\bibinfo {volume} {771}},\ \bibinfo {pages} {51} (\bibinfo {year} {2013})}\BibitemShut {NoStop}%
\bibitem [{\citenamefont {Roca-Maza}\ \emph {et~al.}(2011)\citenamefont {Roca-Maza}, \citenamefont {Centelles}, \citenamefont {Vi\~nas},\ and\ \citenamefont {Warda}}]{Roca-maza}%
  \BibitemOpen
  \bibfield  {author} {\bibinfo {author} {\bibfnamefont {X.}~\bibnamefont {Roca-Maza}}, \bibinfo {author} {\bibfnamefont {M.}~\bibnamefont {Centelles}}, \bibinfo {author} {\bibfnamefont {X.}~\bibnamefont {Vi\~nas}},\ and\ \bibinfo {author} {\bibfnamefont {M.}~\bibnamefont {Warda}},\ }\href {https://doi.org/10.1103/PhysRevLett.106.252501} {\bibfield  {journal} {\bibinfo  {journal} {Phys. Rev. Lett.}\ }\textbf {\bibinfo {volume} {106}},\ \bibinfo {pages} {252501} (\bibinfo {year} {2011})}\BibitemShut {NoStop}%
\bibitem [{\citenamefont {Zenihiro}\ \emph {et~al.}(2010)\citenamefont {Zenihiro} \emph {et~al.}}]{Zenihiro}%
  \BibitemOpen
  \bibfield  {author} {\bibinfo {author} {\bibfnamefont {J.}~\bibnamefont {Zenihiro}} \emph {et~al.},\ }\href {https://doi.org/10.1103/PhysRevC.82.044611} {\bibfield  {journal} {\bibinfo  {journal} {Phys. Rev. C}\ }\textbf {\bibinfo {volume} {82}},\ \bibinfo {pages} {044611} (\bibinfo {year} {2010})}\BibitemShut {NoStop}%
\bibitem [{\citenamefont {Tamii}\ \emph {et~al.}(2011)\citenamefont {Tamii} \emph {et~al.}}]{Tamii2011}%
  \BibitemOpen
  \bibfield  {author} {\bibinfo {author} {\bibfnamefont {A.}~\bibnamefont {Tamii}} \emph {et~al.},\ }\href {https://doi.org/10.1103/PhysRevLett.107.062502} {\bibfield  {journal} {\bibinfo  {journal} {Phys. Rev. Lett.}\ }\textbf {\bibinfo {volume} {107}},\ \bibinfo {pages} {062502} (\bibinfo {year} {2011})}\BibitemShut {NoStop}%
\bibitem [{\citenamefont {Adhikari}\ \emph {et~al.}(2021)\citenamefont {Adhikari} \emph {et~al.}}]{PREX}%
  \BibitemOpen
  \bibfield  {author} {\bibinfo {author} {\bibfnamefont {D.}~\bibnamefont {Adhikari}} \emph {et~al.} (\bibinfo {collaboration} {PREX Collaboration}),\ }\href {https://doi.org/10.1103/PhysRevLett.126.172502} {\bibfield  {journal} {\bibinfo  {journal} {Phys. Rev. Lett.}\ }\textbf {\bibinfo {volume} {126}},\ \bibinfo {pages} {172502} (\bibinfo {year} {2021})}\BibitemShut {NoStop}%
\bibitem [{\citenamefont {Adhikari}\ \emph {et~al.}(2022)\citenamefont {Adhikari} \emph {et~al.}}]{CREX}%
  \BibitemOpen
  \bibfield  {author} {\bibinfo {author} {\bibfnamefont {D.}~\bibnamefont {Adhikari}} \emph {et~al.} (\bibinfo {collaboration} {CREX Collaboration}),\ }\href {https://doi.org/10.1103/PhysRevLett.129.042501} {\bibfield  {journal} {\bibinfo  {journal} {Phys. Rev. Lett.}\ }\textbf {\bibinfo {volume} {129}},\ \bibinfo {pages} {042501} (\bibinfo {year} {2022})}\BibitemShut {NoStop}%
\bibitem [{\citenamefont {Carbone}\ \emph {et~al.}(2010)\citenamefont {Carbone} \emph {et~al.}}]{Carbone2010}%
  \BibitemOpen
  \bibfield  {author} {\bibinfo {author} {\bibfnamefont {A.}~\bibnamefont {Carbone}} \emph {et~al.},\ }\href {https://doi.org/10.1103/PhysRevC.81.041301} {\bibfield  {journal} {\bibinfo  {journal} {Phys. Rev. C}\ }\textbf {\bibinfo {volume} {81}},\ \bibinfo {pages} {041301} (\bibinfo {year} {2010})}\BibitemShut {NoStop}%
\bibitem [{\citenamefont {Berman}\ and\ \citenamefont {Fultz}(1975)}]{Berman_1975}%
  \BibitemOpen
  \bibfield  {author} {\bibinfo {author} {\bibfnamefont {B.~L.}\ \bibnamefont {Berman}}\ and\ \bibinfo {author} {\bibfnamefont {S.~C.}\ \bibnamefont {Fultz}},\ }\href {https://doi.org/10.1103/RevModPhys.47.713} {\bibfield  {journal} {\bibinfo  {journal} {Rev. Mod. Phys.}\ }\textbf {\bibinfo {volume} {47}},\ \bibinfo {pages} {713} (\bibinfo {year} {1975})}\BibitemShut {NoStop}%
\bibitem [{\citenamefont {Tsang}\ \emph {et~al.}(2012)\citenamefont {Tsang} \emph {et~al.}}]{Tsang2012}%
  \BibitemOpen
  \bibfield  {author} {\bibinfo {author} {\bibfnamefont {M.~B.}\ \bibnamefont {Tsang}} \emph {et~al.},\ }\href {https://doi.org/10.1103/PhysRevC.86.015803} {\bibfield  {journal} {\bibinfo  {journal} {Phys. Rev. C}\ }\textbf {\bibinfo {volume} {86}},\ \bibinfo {pages} {015803} (\bibinfo {year} {2012})}\BibitemShut {NoStop}%
\bibitem [{\citenamefont {Tanaka}\ \emph {et~al.}(2017)\citenamefont {Tanaka} \emph {et~al.}}]{kilonova}%
  \BibitemOpen
  \bibfield  {author} {\bibinfo {author} {\bibfnamefont {M.}~\bibnamefont {Tanaka}} \emph {et~al.},\ }\href {https://doi.org/10.1093/pasj/psx121} {\bibfield  {journal} {\bibinfo  {journal} {Publications of the Astronomical Society of Japan}\ }\textbf {\bibinfo {volume} {69}},\ \bibinfo {pages} {102} (\bibinfo {year} {2017})}\BibitemShut {NoStop}%
\bibitem [{\citenamefont {Abbott}\ \emph {et~al.}(2017)\citenamefont {Abbott} \emph {et~al.}}]{GW170817}%
  \BibitemOpen
  \bibfield  {author} {\bibinfo {author} {\bibfnamefont {B.~P.}\ \bibnamefont {Abbott}} \emph {et~al.} (\bibinfo {collaboration} {LIGO Scientific Collaboration and Virgo Collaboration}),\ }\href {https://doi.org/10.1103/PhysRevLett.119.161101} {\bibfield  {journal} {\bibinfo  {journal} {Phys. Rev. Lett.}\ }\textbf {\bibinfo {volume} {119}},\ \bibinfo {pages} {161101} (\bibinfo {year} {2017})}\BibitemShut {NoStop}%
\bibitem [{\citenamefont {Evans}\ \emph {et~al.}(2017)\citenamefont {Evans} \emph {et~al.}}]{Evans}%
  \BibitemOpen
  \bibfield  {author} {\bibinfo {author} {\bibfnamefont {P.~A.}\ \bibnamefont {Evans}} \emph {et~al.},\ }\href {https://doi.org/10.1126/science.aap9580} {\bibfield  {journal} {\bibinfo  {journal} {Science}\ }\textbf {\bibinfo {volume} {358}},\ \bibinfo {pages} {1565} (\bibinfo {year} {2017})}\BibitemShut {NoStop}%
\bibitem [{\citenamefont {Metzger}\ \emph {et~al.}(2018)\citenamefont {Metzger}, \citenamefont {Thompson},\ and\ \citenamefont {Quataert}}]{Metzger_2018}%
  \BibitemOpen
  \bibfield  {author} {\bibinfo {author} {\bibfnamefont {B.~D.}\ \bibnamefont {Metzger}}, \bibinfo {author} {\bibfnamefont {T.~A.}\ \bibnamefont {Thompson}},\ and\ \bibinfo {author} {\bibfnamefont {E.}~\bibnamefont {Quataert}},\ }\href {https://doi.org/10.3847/1538-4357/aab095} {\bibfield  {journal} {\bibinfo  {journal} {Astrophys. J.}\ }\textbf {\bibinfo {volume} {856}},\ \bibinfo {pages} {101} (\bibinfo {year} {2018})}\BibitemShut {NoStop}%
\bibitem [{\citenamefont {Tachibana}\ \emph {et~al.}(2025)\citenamefont {Tachibana}, \citenamefont {Hagino}, \citenamefont {Yoshida},\ and\ \citenamefont {Zhao}}]{Tachibana2025}%
  \BibitemOpen
  \bibfield  {author} {\bibinfo {author} {\bibfnamefont {T.}~\bibnamefont {Tachibana}}, \bibinfo {author} {\bibfnamefont {K.}~\bibnamefont {Hagino}}, \bibinfo {author} {\bibfnamefont {K.}~\bibnamefont {Yoshida}},\ and\ \bibinfo {author} {\bibfnamefont {Q.}~\bibnamefont {Zhao}},\ }\href {https://doi.org/10.1103/8sv1-t27l} {\bibfield  {journal} {\bibinfo  {journal} {Phys. Rev. C}\ }\textbf {\bibinfo {volume} {112}},\ \bibinfo {pages} {065806} (\bibinfo {year} {2025})}\BibitemShut {NoStop}%
\bibitem [{\citenamefont {Tan}\ \emph {et~al.}(2020)\citenamefont {Tan}, \citenamefont {Khoa},\ and\ \citenamefont {Loan}}]{Khoa_2020}%
  \BibitemOpen
  \bibfield  {author} {\bibinfo {author} {\bibfnamefont {N.~H.}\ \bibnamefont {Tan}}, \bibinfo {author} {\bibfnamefont {D.~T.}\ \bibnamefont {Khoa}},\ and\ \bibinfo {author} {\bibfnamefont {D.~T.}\ \bibnamefont {Loan}},\ }\href {https://doi.org/10.1103/PhysRevC.102.045809} {\bibfield  {journal} {\bibinfo  {journal} {Phys. Rev. C}\ }\textbf {\bibinfo {volume} {102}},\ \bibinfo {pages} {045809} (\bibinfo {year} {2020})}\BibitemShut {NoStop}%
\bibitem [{\citenamefont {Khoa}\ \emph {et~al.}(2022)\citenamefont {Khoa}, \citenamefont {Tan},\ and\ \citenamefont {Khoa}}]{Khoa_2022}%
  \BibitemOpen
  \bibfield  {author} {\bibinfo {author} {\bibfnamefont {N.~H.~D.}\ \bibnamefont {Khoa}}, \bibinfo {author} {\bibfnamefont {N.~H.}\ \bibnamefont {Tan}},\ and\ \bibinfo {author} {\bibfnamefont {D.~T.}\ \bibnamefont {Khoa}},\ }\href {https://doi.org/10.1103/PhysRevC.105.065802} {\bibfield  {journal} {\bibinfo  {journal} {Phys. Rev. C}\ }\textbf {\bibinfo {volume} {105}},\ \bibinfo {pages} {065802} (\bibinfo {year} {2022})}\BibitemShut {NoStop}%
\bibitem [{\citenamefont {Gadea}\ \emph {et~al.}(2005)\citenamefont {Gadea} \emph {et~al.}}]{Gadea2005}%
  \BibitemOpen
  \bibfield  {author} {\bibinfo {author} {\bibfnamefont {A.}~\bibnamefont {Gadea}} \emph {et~al.},\ }\href {https://doi.org/https://doi.org/10.1016/j.physletb.2005.05.073} {\bibfield  {journal} {\bibinfo  {journal} {Physics Letters B}\ }\textbf {\bibinfo {volume} {619}},\ \bibinfo {pages} {88} (\bibinfo {year} {2005})}\BibitemShut {NoStop}%
\bibitem [{\citenamefont {Ur}\ \emph {et~al.}(1998)\citenamefont {Ur} \emph {et~al.}}]{Ur1998}%
  \BibitemOpen
  \bibfield  {author} {\bibinfo {author} {\bibfnamefont {C.~A.}\ \bibnamefont {Ur}} \emph {et~al.},\ }\href {https://doi.org/10.1103/PhysRevC.58.3163} {\bibfield  {journal} {\bibinfo  {journal} {Phys. Rev. C}\ }\textbf {\bibinfo {volume} {58}},\ \bibinfo {pages} {3163} (\bibinfo {year} {1998})}\BibitemShut {NoStop}%
\bibitem [{\citenamefont {Geesaman}\ \emph {et~al.}(1979)\citenamefont {Geesaman}, \citenamefont {McGrath}, \citenamefont {No\'e},\ and\ \citenamefont {Malmin}}]{Geesaman1979}%
  \BibitemOpen
  \bibfield  {author} {\bibinfo {author} {\bibfnamefont {D.~F.}\ \bibnamefont {Geesaman}}, \bibinfo {author} {\bibfnamefont {R.~L.}\ \bibnamefont {McGrath}}, \bibinfo {author} {\bibfnamefont {J.~W.}\ \bibnamefont {No\'e}},\ and\ \bibinfo {author} {\bibfnamefont {R.~E.}\ \bibnamefont {Malmin}},\ }\href {https://doi.org/10.1103/PhysRevC.19.1938} {\bibfield  {journal} {\bibinfo  {journal} {Phys. Rev. C}\ }\textbf {\bibinfo {volume} {19}},\ \bibinfo {pages} {1938} (\bibinfo {year} {1979})}\BibitemShut {NoStop}%
\bibitem [{\citenamefont {Liang}\ \emph {et~al.}(2018)\citenamefont {Liang}, \citenamefont {Sagawa}, \citenamefont {Sasano}, \citenamefont {Suzuki},\ and\ \citenamefont {Honma}}]{Liang2018}%
  \BibitemOpen
  \bibfield  {author} {\bibinfo {author} {\bibfnamefont {H.~Z.}\ \bibnamefont {Liang}}, \bibinfo {author} {\bibfnamefont {H.}~\bibnamefont {Sagawa}}, \bibinfo {author} {\bibfnamefont {M.}~\bibnamefont {Sasano}}, \bibinfo {author} {\bibfnamefont {T.}~\bibnamefont {Suzuki}},\ and\ \bibinfo {author} {\bibfnamefont {M.}~\bibnamefont {Honma}},\ }\href {https://doi.org/10.1103/PhysRevC.98.014311} {\bibfield  {journal} {\bibinfo  {journal} {Phys. Rev. C}\ }\textbf {\bibinfo {volume} {98}},\ \bibinfo {pages} {014311} (\bibinfo {year} {2018})}\BibitemShut {NoStop}%
\bibitem [{\citenamefont {Zhao}\ \emph {et~al.}(2022)\citenamefont {Zhao}, \citenamefont {Ren}, \citenamefont {Zhao},\ and\ \citenamefont {Meng}}]{PCF-PK1}%
  \BibitemOpen
  \bibfield  {author} {\bibinfo {author} {\bibfnamefont {Q.}~\bibnamefont {Zhao}}, \bibinfo {author} {\bibfnamefont {Z.}~\bibnamefont {Ren}}, \bibinfo {author} {\bibfnamefont {P.}~\bibnamefont {Zhao}},\ and\ \bibinfo {author} {\bibfnamefont {J.}~\bibnamefont {Meng}},\ }\href {https://doi.org/10.1103/PhysRevC.106.034315} {\bibfield  {journal} {\bibinfo  {journal} {Phys. Rev. C}\ }\textbf {\bibinfo {volume} {106}},\ \bibinfo {pages} {034315} (\bibinfo {year} {2022})}\BibitemShut {NoStop}%
\bibitem [{\citenamefont {B\"urvenich}\ \emph {et~al.}(2002)\citenamefont {B\"urvenich}, \citenamefont {Madland}, \citenamefont {Maruhn},\ and\ \citenamefont {Reinhard}}]{PC-F1}%
  \BibitemOpen
  \bibfield  {author} {\bibinfo {author} {\bibfnamefont {T.}~\bibnamefont {B\"urvenich}}, \bibinfo {author} {\bibfnamefont {D.~G.}\ \bibnamefont {Madland}}, \bibinfo {author} {\bibfnamefont {J.~A.}\ \bibnamefont {Maruhn}},\ and\ \bibinfo {author} {\bibfnamefont {P.-G.}\ \bibnamefont {Reinhard}},\ }\href {https://doi.org/10.1103/PhysRevC.65.044308} {\bibfield  {journal} {\bibinfo  {journal} {Phys. Rev. C}\ }\textbf {\bibinfo {volume} {65}},\ \bibinfo {pages} {044308} (\bibinfo {year} {2002})}\BibitemShut {NoStop}%
\bibitem [{\citenamefont {Zhao}\ \emph {et~al.}(2010)\citenamefont {Zhao}, \citenamefont {Li}, \citenamefont {Yao},\ and\ \citenamefont {Meng}}]{PC-PK1}%
  \BibitemOpen
  \bibfield  {author} {\bibinfo {author} {\bibfnamefont {P.~W.}\ \bibnamefont {Zhao}}, \bibinfo {author} {\bibfnamefont {Z.~P.}\ \bibnamefont {Li}}, \bibinfo {author} {\bibfnamefont {J.~M.}\ \bibnamefont {Yao}},\ and\ \bibinfo {author} {\bibfnamefont {J.}~\bibnamefont {Meng}},\ }\href {https://doi.org/10.1103/PhysRevC.82.054319} {\bibfield  {journal} {\bibinfo  {journal} {Phys. Rev. C}\ }\textbf {\bibinfo {volume} {82}},\ \bibinfo {pages} {054319} (\bibinfo {year} {2010})}\BibitemShut {NoStop}%
\bibitem [{\citenamefont {Ring}\ and\ \citenamefont {Schuck}(1980)}]{RingSchuck}%
  \BibitemOpen
  \bibfield  {author} {\bibinfo {author} {\bibfnamefont {P.}~\bibnamefont {Ring}}\ and\ \bibinfo {author} {\bibfnamefont {P.}~\bibnamefont {Schuck}},\ }\href@noop {} {\emph {\bibinfo {title} {The Nuclear Many-Body Problem}}}\ (\bibinfo  {publisher} {Springer Verlag Berlin-Heidelberg},\ \bibinfo {year} {1980})\BibitemShut {NoStop}%
\bibitem [{\citenamefont {Tian}\ \emph {et~al.}(2009)\citenamefont {Tian}, \citenamefont {Ma},\ and\ \citenamefont {Ring}}]{Tian2009}%
  \BibitemOpen
  \bibfield  {author} {\bibinfo {author} {\bibfnamefont {Y.}~\bibnamefont {Tian}}, \bibinfo {author} {\bibfnamefont {Z.}~\bibnamefont {Ma}},\ and\ \bibinfo {author} {\bibfnamefont {P.}~\bibnamefont {Ring}},\ }\href {https://doi.org/https://doi.org/10.1016/j.physletb.2009.04.067} {\bibfield  {journal} {\bibinfo  {journal} {Physics Letters B}\ }\textbf {\bibinfo {volume} {676}},\ \bibinfo {pages} {44} (\bibinfo {year} {2009})}\BibitemShut {NoStop}%
\bibitem [{\citenamefont {Szyma{\'n}ski}(1983)}]{szymański1983fast}%
  \BibitemOpen
  \bibfield  {author} {\bibinfo {author} {\bibfnamefont {Z.}~\bibnamefont {Szyma{\'n}ski}},\ }\href@noop {} {\emph {\bibinfo {title} {Fast Nuclear Rotation}}},\ Oxford science publications\ (\bibinfo  {publisher} {Clarendon Press},\ \bibinfo {year} {1983})\BibitemShut {NoStop}%
\bibitem [{\citenamefont {Inakura}\ and\ \citenamefont {Nakada}(2015)}]{Inakura2015}%
  \BibitemOpen
  \bibfield  {author} {\bibinfo {author} {\bibfnamefont {T.}~\bibnamefont {Inakura}}\ and\ \bibinfo {author} {\bibfnamefont {H.}~\bibnamefont {Nakada}},\ }\href {https://doi.org/10.1103/PhysRevC.92.064302} {\bibfield  {journal} {\bibinfo  {journal} {Phys. Rev. C}\ }\textbf {\bibinfo {volume} {92}},\ \bibinfo {pages} {064302} (\bibinfo {year} {2015})}\BibitemShut {NoStop}%
\bibitem [{\citenamefont {Koszor\'{u}s}\ \emph {et~al.}(2021)\citenamefont {Koszor\'{u}s} \emph {et~al.}}]{Koszorus2021}%
  \BibitemOpen
  \bibfield  {author} {\bibinfo {author} {\bibfnamefont {A.}~\bibnamefont {Koszor\'{u}s}} \emph {et~al.},\ }\href {https://doi.org/https://doi.org/10.1016/j.physletb.2021.136439} {\bibfield  {journal} {\bibinfo  {journal} {Physics Letters B}\ }\textbf {\bibinfo {volume} {819}},\ \bibinfo {pages} {136439} (\bibinfo {year} {2021})}\BibitemShut {NoStop}%
\bibitem [{\citenamefont {Vernon}\ \emph {et~al.}(2025)\citenamefont {Vernon} \emph {et~al.}}]{Vernon2025}%
  \BibitemOpen
  \bibfield  {author} {\bibinfo {author} {\bibfnamefont {A.~R.}\ \bibnamefont {Vernon}} \emph {et~al.},\ }\href {https://doi.org/10.1103/m35x-mw7z} {\bibfield  {journal} {\bibinfo  {journal} {Phys. Rev. Lett.}\ }\textbf {\bibinfo {volume} {134}},\ \bibinfo {pages} {252501} (\bibinfo {year} {2025})}\BibitemShut {NoStop}%
\end{thebibliography}%
\end{document}